\documentstyle[11pt,newpasp]{article}

\begin{document}

\title{Luminous red mergers in the $z=0.83$ cluster MS\,1054--03: Direct
evidence for hierarchical formation of massive early-type galaxies}

\author{Pieter van Dokkum\altaffilmark{1,2} and Marijn Franx}
\affil{Leiden Observatory, P.\ O.\ Box 9513, 2300 RA Leiden, The Netherlands}
\author{Daniel Fabricant}
\affil{Harvard-Smithsonian Center for Astrophysics, 60 Garden Street,
Cambridge, MA 02318}
\author{Daniel Kelson}
\affil{Department of Terrestrial Magnetism, 5241 Broad Branch Road,
NW, Washington, DC, 20015-1305}
\author{Garth Illingworth}
\affil{University of California, Santa Cruz, CA 95064}


\altaffiltext{1}{Kapteyn Astronomical Institute, P.\ O.\ Box 800,
9700 AV Groningen, The Netherlands}
\altaffiltext{2}{Present address: California Institute of Technology,
Pasadena, CA 91125}



\begin{abstract}
We present results from a morphological study of the distant X-ray
cluster MS\,1054--03 at $z=0.83$. We have obtained a large, two color
HST WFPC2 mosaic of this cluster, and measured redshifts of 186
galaxies in the MS\,1054--03 field with the 10m Keck telescope.  Of 81
spectroscopically confirmed cluster galaxies observerd with HST, 13
are merger systems. Most of these mergers will likely evolve into
luminous elliptical galaxies, and some may evolve into S0 galaxies. If
the galaxy population in MS\,1054--03 is typical for its redshift up
to $\sim 50$\,\% of ellipticals may have formed in mergers at
$z<1$. The mergers are generally red and have no detected
[O{\sc ii}]\,3727\,\AA{} emission.
This result is consistent with the old stellar ages of
ellipticals inferred from other studies. The mergers are
preferentially found towards the outskirts of the cluster, indicating
they probably occur in infalling clumps. A significant overabundance of
close pairs of red galaxies detected in the outskirts of MS\,1054--03
confirms the large number of interacting galaxies in this cluster.
\end{abstract}




\section{Introduction}
The formation epoch of early-type galaxies provides a strong test
for galaxy formation models. Traditional models assume early-type
galaxies formed in a ``monolithic'' collapse at very high redshift,
followed by a smooth and regular dimming of the stellar light
(e.g., Searle et al.\ 1973). In contrast,
currently popular models for galaxy formation in CDM cosmologies
predict that early-type
galaxies were formed in many generations of mergers, and many
ellipticals experienced their last major merger at $z < 1$ (e.g.,
Baugh, Cole, \& Frenk 1996).

Early-type galaxies are easily studied in clusters, and most of what
we have learned about the formation and evolution of early-types
has come from studies of rich clusters at $0<z<1$ (e.g.,
Dressler et al.\ 1997). Strong constraints
on the ages of the stars in early-type galaxies have come from studies
of the evolution of the color-magnitude relation (e.g., Ellis et al.\ 1997,
Stanford et al.\ 1998) and the Fundamental Plane (e.g., van Dokkum
et al.\ 1998). These studies are all in remarkable agreement:
most of the stars in early-type galaxies appear to have formed
at high redshift ($z>2$), and there is very little cluster-to-cluster
scatter in their properties.

However, as pointed out by, e.g., Kauffmann (1996) the time of
assembly of massive galaxies may be much more recent than the
mean age of their stars. The FP and the CM relation do not provide
constraints on the assembly time of early-type galaxies, unless
some assumption is made regarding the amount of star formation
during and prior to the mergers.

The relevance of merging can be constrained by other means.
In particular, by studying large samples of distant
galaxies the fraction of the galaxy population
which is (at a given epoch) involved in a merger can
be determined (e.g., Le Fevre et al.\ 1999).
Here, we report on the merger fraction in
the cluster MS\,1054--03 at $z=0.83$. We have obtained a large
HST WFPC2 mosaic of the cluster, and combined this with
extensive spectroscopy with the Keck telescope. The
sample consists of 81 confirmed cluster members observed
with HST. The results of this study are presented in full in
van Dokkum et al.\ (1999).

\section{Mergers in MS\,1054--03}

The most surprising result of our survey is the high fraction
of galaxies classified as ``merger/peculiar''. We classified
17\,\% as mergers, compared to 22\,\% ellipticals, 22\,\% S0s,
and 39\,\% spirals. Examples of the mergers are shown in
Figure 1. We emphasize that all classified galaxies,
including the mergers, are spectroscopically
confirmed cluster galaxies.

\begin{figure*}
\includegraphics{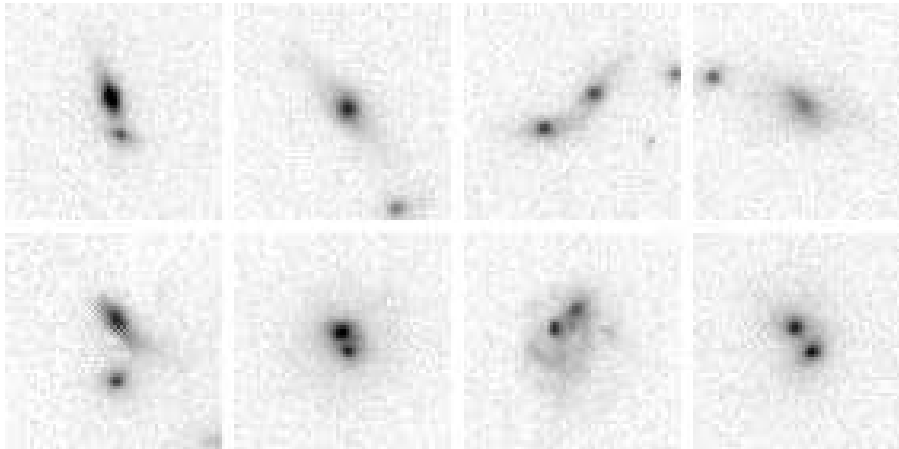}
\vspace{6.1cm}
\caption{Examples of mergers in MS\,1054--03 at $z=0.83$. All galaxies
are spectroscopically confirmed cluster members. These galaxies
are among the most luminous in the cluster. Most of them will probably
evolve into luminous ($\sim 2 L_*$) ellipticals within $\sim 1$\,Gyr.}
\label{mergers.fig}
\end{figure*}

The mergers are very luminous:
the 13 mergers have
a median luminosity $M_B^T \approx -22$ ($\sim 2 L_*$ at $z=0.83$),
and five of the sixteen most luminous
cluster galaxies were classified as mergers.
The majority of the mergers will probably evolve into elliptical
galaxies. The merger fraction in MS\,1054 is comparable to
the elliptical fraction: the number of ellipticals
``in formation'' is similar to the number of ellipticals already
formed. Assuming the galaxy population in MS\,1054--03 is typical
for its redshift, this implies up to $\sim 50$\,\% of
ellipticals formed in mergers at $z<1$.

The high merger fraction in MS\,1054--03 may seem surprising given
the high velocity dispersion of the cluster ($\approx 1170$\,km\,s$^{-1}$,
Tran et al.\ 1999). The progenitors of the mergers must have had
much lower relative velocities, indicating they were part of
cold subclumps, which may in turn be merging with the cluster.
The spatial distribution of the mergers supports this idea:
they are preferentially located towards the outskirts of the cluster
(see van Dokkum et al.\ 1999).

\section{Close pairs}

The merger fraction can be determined independently by
examining the number of close pairs. Figure 2 shows the
average galaxy density around red galaxies in our HST mosaic
of MS\,1054--03, excluding the central $400 h_{50}^{-1}$\,kpc of
the cluster. The number of galaxies in each bin is weighted
by the area; therefore, a flat distribution would show that
the galaxies in the outer parts of the cluster are distributed
uniformly.

There is a pronounced peak at separations $<20 h_{50}^{-1}$\,kpc.
This (preliminary) result confirms
the large number of interacting galaxies in this cluster, and is
completely independent from the morphological classifications.

\begin{figure*}
\includegraphics{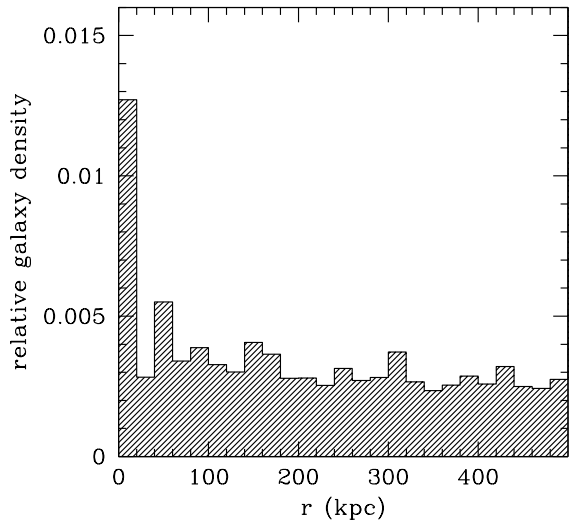}
\vspace{6.9cm}
\caption{The average galaxy density around red galaxies in our HST
mosaic of MS\,1054--03, excluding the core of the cluster.
There is a pronounced peak at separations
$<20 h_{50}^{-1}$\,kpc (or $2\arcsec$), demonstrating the presence
of a large number of close pairs in the outskirts of MS\,1054--03.}
\label{pairs.fig}
\end{figure*}

\section{Discussion}
It may seem difficult to reconcile our results
with the high formation redshifts inferred from studies of the
luminosity and color evolution of early-type galaxies
to $z=1$ (e.g., Stanford et al.\ 1998, van Dokkum et al.\ 1998).
However, this apparent conflict is at least partially
solved by the observation
that the merging galaxies in MS\,1054--03 are red, and seem to have
evolved stellar populations. Their colors indicate that the
bulk of their stars was formed at $z>1.7$
(see van Dokkum et al.\ 1999).
The mergers are typically bulge dominated and have no detected
[O{\sc ii}]\,3727\,\AA{} emission; they appear to be
mergers between E/S0s or early-type spirals.
The available evidence suggests that the stars in massive
early-types were formed at high redshift ($z>2$),
whereas the assembly
of the galaxies themselves continued to much lower redshift
($z<1$). Our findings are strong evidence against galaxy
formation in a monolithic collapse at high redshift,
and in qualitative agreement with hierarchical galaxy
formation models.

This study demonstrates that the combination of
large field imaging with HST and deep spectroscopy from the ground
can show directly how galaxy formation proceeded.
The large field was essential, since the mergers
are preferentially located in the outskirts of the cluster.
Similar observations of high redshift clusters and the field
would be valuable to test whether the results for
MS\,1054--03 are typical.


\begin{references}
\reference Baugh,  C. M., Cole, S., \& Frenk, C. S. 1996, \mnras,
	283, 1361
\reference Dressler, A., et al. 1997,  \apj, 490, 577
\reference Ellis, R. S., et al. 1997, \apj, 483, 582
\reference Kauffmann, G. 1996, \mnras, 281, 487
\reference Le Fevre, O., et al. 1999, \mnras, accepted
	(astro-ph/9909211)
\reference Searle, L., Sargent, W. L. W., \& Bagnuolo, W. G. 1973,
	\apj, 179, 427
\reference Stanford, S. A., Eisenhardt, P. R., \& Dickinson, M.
        1998, \apj, 492, 461
\reference Tran, K-V. H., et al. 1999, \apj, 522, 39
\reference van Dokkum, P. G., Franx, M., Kelson, D. D.,
        Illingworth, G. D. 1998, \apj, 504, L17
\reference van Dokkum, P. G., Franx, M., Fabricant, D., Kelson, D. D.,
	Illingworth, G. D. 1999, \apj, 520, L95
\end{references}
\end{document}